\documentstyle[aps,amsfonts]{revtex}

\def\R{{\Bbb R}}
\def\Ham{{\cal H}_{\rm cl}}
\def\H{\hat H}
\def\J{\hat J}
\def\V{\hat V}
\def\Jzero{{\hat J}_0}
\def\Hzero{{\hat H}_0}
\def\Jxi{{\hat J}_\xi}
\def\Hxi{{\hat H}_\xi}
\def\CE2{CE_{\cal H}}

\newtheorem{theorem}{Theorem}
\newtheorem{lemma}{Lemma}
\def\ds{\displaystyle}
\def\endproof{{\ \vbox{\hrule\hbox{%
   \vrule height1.5ex\hskip1.1ex\vrule}\hrule
  }}\par}

\begin{document}

\title{
{\normalsize gr-qc/9807028}
%\hfill{\normalsize DRAFT, \today}
\hfill{\normalsize 13 July 1998}
\\[3mm]
  {\bf A NEW LOOK AT THE ASHTEKAR-MAGNON ENERGY CONDITION}
}

\author{Jim Fischer \& Tevian Dray}
\address{Department of Mathematics, Oregon State University,
		Corvallis, OR  97331, USA}

\maketitle

%\pacs{04.20.Cv, 04.20.Me, 11.30.-j, 02.40.Hw}

\begin{abstract}
In 1975, Ashtekar and Magnon~\cite{AM} showed that an energy condition selects
a unique quantization procedure for certain observers in general, curved
spacetimes.  We generalize this result in two important ways, by eliminating
the need to assume a particular form for the (quantum) Hamiltonian, and by
considering the surprisingly nontrivial extension to nonminimal coupling.
\end{abstract}

\section{Introduction}

Ashtekar and Magnon~\cite{AM} were among the first to consider quantum field
theory as seen by observers who were {\it not} static or stationary.
Remarkably, they were able to give a quantization procedure for the scalar
field for {\it any} family of hypersurface orthogonal observers in a curved
spacetime.  Their procedure is based on a single, natural condition: The
classical and quantum energies should agree.  However, for non-Killing
observers, it is not obvious how to define either of these energies.  Ashtekar
and Magnon choose to use the stress-energy tensor of the scalar field for the
classical energy, and to define the quantum energy in terms of a particular
choice of quantum Hamiltonian operator.

We extend their work in two important ways.  First of all, we show that the
construction itself fully determines the Hamiltonian operator, which therefore
does not need to be specified in advance.  Second, we show that the basic
result holds for {\it any} choice of the classical energy satisfying certain
simple properties.  Not surprisingly, for the case of minimal coupling
($\xi=0$) considered by Ashtekar and Magnon, if we use the stress-energy
tensor to define the classical energy, then we recover not only their complex
structure but also their quantum Hamiltonian.  However, when $\xi\ne0$, this
approach runs into a serious problem: The resulting Hamiltonian and complex
structure do not reduce to the known answers for static observers.  We show
how this problem can be resolved by using the classical Hamiltonian to define
the classical energy rather than the stress-energy tensor.

After setting up our formalism in Section~\ref{Math}, we summarize the work of
Ashtekar and Magnon in Section~\ref{Energy} and present our generalization in
Section~\ref{Result}.  Section~\ref{Appl} shows how to recover Ashtekar and
Magnon's result for $\xi=0$, as well as considering the case $\xi\ne0$.
Finally, in Section~\ref{Disc} we discuss our results.

%%%%%%%%%%%%%%%%%%%%%%%%%%%%%%%%%%%%%%%%%%%%%%%%%%%%%%%%%%%%%%%%%%%%%%%%%%%%%%
% Chapter 2

\section{Mathematical Preliminaries}
\label{Math}

Let $(M,g_{ab})$ be a globally hyperbolic spacetime with associated
Levi-Civita connection $\nabla$.  The action ${\cal S}$ for a scalar field
$\Phi$ on $M$ is given by
\begin{eqnarray}
{\cal S} &=& \int_M {\cal{L}} \,\sqrt{-g}\,d^nx
\label{action}\\
{\cal L} &=& -\frac{1}{2}\left( g^{ab}\nabla_a\Phi\nabla_b\Phi
				+ (m^2 + \xi R)\Phi^2 \right)
\label{lagrange}
\end{eqnarray}
The Klein-Gordon equation, obtained by varying the action $\cal S$ with
respect to $\Phi$, is
\begin{equation}
g^{ab}\nabla_a\nabla_b\Phi - (m^2 + \xi R)\Phi = 0
\label{KG}
\end{equation}
Let $V$ be the space of smooth, real-valued solutions of (\ref{KG}) which
have compact support on any (and hence every) Cauchy surface.  Ashtekar and
Magnon suggested that, as a real vector space, the one-particle Hilbert space
${\cal{H}}$ should be a copy of $V$.

Introduce coordinates $(t=x_0,\dots,x_{n-1})$ on $M$ so that the hypersurfaces
$\{t={\rm const}\}$ are Cauchy surfaces.  We assume throughout that the vector
field $t^a\nabla_a=\partial_t$ is hypersurface orthogonal.  The standard $3+1$
formalism leads to a decomposition of the metric $g_{ab}$ in terms of its
pullback $h_{ij}$ to $\Sigma$ and the lapse function $N=t^at_a$; the shift is
zero.  We denote the Levi-Civita connection on $(\Sigma,h_{ij})$ by $D_i$.

On $V$ we have the (nondegenerate) symplectic structure
\begin{equation}
\Omega(\Phi,\Psi)
  = \int_\Sigma \left(\Psi \nabla_a\Phi - \Phi\nabla_a\Psi \right)\, n^a d\Sigma
\label{symstruc}
\end{equation}
where $\Sigma$ is any Cauchy hypersurface, $n^a$ is the future pointing, unit
normal vector field to $\Sigma$ and $d\Sigma= \sqrt{h}\, d^{n-1}x$ is the
volume element on $\Sigma$ induced by the inclusion map.  Let $J$ be any
complex structure on $V$, that is a linear map from $V$ to itself satisfying
\begin{equation}
J^2 = -1
\label{struct}
\end{equation}
which allows us to view $V$ as a complex vector space.  We will also assume
that $J$ is compatible in the sense that
\begin{eqnarray}
\Omega(\Phi,J\Psi) &\geq& 0 \label{com1}\\
\Omega(J\Phi,J\Psi) &=& \Omega(\Phi, \Psi) \label{com2}
\end{eqnarray}
for any $\Phi,\Psi\in V$.  As discussed in~\cite{AM}, the $*$-algebra approach
leads naturally to the inner product
\begin{equation}
\left<\Phi,\Psi\right>
  = \frac{1}{2}\Omega(\Phi, J\Psi) + \frac{i}{2}\Omega(\Phi, \Psi)
\label{innerproduct}
\end{equation}
which is Hermitian under the above assumptions.  A candidate for the
one-particle Hilbert space ${\cal H}$ is then the Cauchy completion of
$(V,J,\left<\_,\_\right>)$, so that the problem of identifying the
one-particle Hilbert space of states is reduced to that of choosing a suitable
complex structure $J$ on $V$.

Solutions $\Phi\in V$ are completely determined by their initial data,
\footnote{Ashtekar and Magnon omit the factor of $\sqrt{h}$ from $\pi$.}
so that $V$ is isomorphic to the vector space $\V$ of pairs of smooth,
real-valued functions on $\Sigma$ which have compact support.  The isomorphic
image of $\Phi$ is then
\begin{equation}
\hat\Phi
  = \pmatrix{\varphi\cr \pi}
  = \pmatrix{\Phi\vert_\Sigma\cr \sqrt{h}\,n^a\nabla_a\Phi\vert_\Sigma}
\label{data}
\end{equation}
We write $\tau = C^\infty_0(\Sigma,\R)$, with inner product
\begin{equation}
(f,g) = \int_{\Sigma} fg \,d^{n-1}x\\
\end{equation}
for $f,g\in\V$; note that $\V=\tau\oplus\tau$.

We conclude this section with some results about symmetric operators.  Any
linear operator $Q$ on $V$ can be represented as a $2\times2$ matrix $\hat Q$
whose elements are linear operators on $\tau$.  In particular, we write
\begin{eqnarray}
\J &=& \pmatrix{A& B\cr C& D}\\
\noalign{\medskip}
\H &=& \pmatrix{W& X\cr Y& Z}
\end{eqnarray}
We define $Q$ to be {\it symmetric} on $V$ if
\begin{equation}
\left<\Phi,Q\Psi\right> = \left<Q\Phi,\Psi\right>
\label{symmetry}
\end{equation}
for all $\Phi,\Psi\in V$; $Q$ is {\it antisymmetric} if a relative factor of
$-1$ is inserted in (\ref{symmetry}).

\begin{lemma}
\label{symtheo2}
Suppose that the linear operator $Q$ on $V$ satisfies
\begin{equation}
\Omega(\Phi,Q\Phi) = 0
\label{Asymm}
\end{equation}
Then $Q$ is symmetric.
\end{lemma}
{\bf Proof:}
This follows immediately since (\ref{Asymm}) implies that the expectation
value of $Q$ is always real.{\mbox{\endproof}}

\begin{lemma}
\label{JHHJ0}
Let $Q$ be a symmetric operator defined on ${\cal H}$.  Then $[Q,J] = 0$.
\end{lemma}
{\bf Proof:} Splitting (\ref{symmetry}) into its real and imaginary parts, we
obtain
\begin{eqnarray}
\Omega(\Phi, JQ\Psi) &=& \Omega(Q\Phi, J\Psi) \label{realpart}\\
\Omega(\Phi, Q\Psi) &=& \Omega(Q\Phi, \Psi) \label{imajpart}
\end{eqnarray}
Using (\ref{realpart}) and (\ref{imajpart}) we obtain:
\begin{eqnarray}
\Omega(\Phi,(JQ - QJ)\Psi) &=& \Omega(\Phi,JQ\Psi) - \Omega(\Phi,QJ\Psi) \\
 &=& \Omega(Q\Phi, J\Psi) - \Omega(Q\Phi, J\Psi) \\
 &=& 0.
\end{eqnarray}
Since $\Phi$ and $\Psi$ are arbitrary (and $\Omega$ is non-degenerate) we
conclude that $JQ -QJ = 0.
${\mbox{\endproof}}

The significance of Lemma \ref{JHHJ0} comes from the fact that the Hamiltonian
operator $H$ should be self-adjoint and hence symmetric.  The total derivative
of $J$ is given by
\begin{equation}
\partial_t J + J [H,J] = 0
\label{totalderiv}
\end{equation}
If $[H,J] = 0$, (\ref{totalderiv}) reduces to $\partial_t J = 0$.  Thus, if
$H$ is self-adjoint, the time derivative of $J$ measures the amount of
particle creation.

Setting $Q=H$, the conditions (15) and (16) for the symmetry of $H$ become
\begin{eqnarray}
X  &=&  -X^{\dag}\label{firsteq}\\
Y &=& -Y^{\dag}\label{seceq}\\
W &=&  Z^{\dag}\label{thirdeq}
\end{eqnarray}
and
\begin{eqnarray}
(AX + BZ) &=& (AX + BZ)^{\dag}\label{fourtheq}\\
(CW + DY) &=& (CW + DY)^{\dag}\label{fiftheq}\\
(AW + BY) &=& -(CX + DZ)^{\dag}\label{sixeq}
\end{eqnarray}
respectively.  But an immediate consequence of Lemma 2 is that
(\ref{realpart}) and (\ref{imajpart}) are equivalent.  Thus, $H$ is symmetric
if and only if (\ref{firsteq})--(\ref{thirdeq}) are satisfied, and this is
equivalent to (\ref{fourtheq})--(\ref{sixeq}) being satisfied.  Furthermore,
these latter equations are precisely the condition for $JH$ to be
antisymmetric, so that we have the further result
\begin{lemma}
$Q$ is symmetric if and only if $JQ$ is antisymmetric.
\end{lemma}

%%%%%%%%%%%%%%%%%%%%%%%%%%%%%%%%%%%%%%%%%%%%%%%%%%%%%%%%%%%%%%%%%%%%%%%%%%%%%%
% Chapter 3

\section{The Ashtekar-Magnon Energy Condition}
\label{Energy}

The essential ingredient in the result of Ashtekar and Magnon~\cite{AM} is the
{\it energy condition}.  Given a Cauchy hypersurface $\Sigma$, one can define
the classical energy and the quantum energy of a scalar field with respect to
those observers orthogonal to $\Sigma$.  The energy condition says that these
energies should be equal.  Ashtekar and Magnon showed that there is a unique
complex structure $J$ on $\Sigma$ such that the energy condition is satisfied.
Using the results of Section~\ref{Math}, they have thus shown that the energy
condition selects a unique quantization procedure.

Ashtekar and Magnon define the classical energy of the scalar field with
respect to $\Sigma$ (and the choice of scale implicit in $t^a$) to be
\begin{equation}
CE_T = \int_\Sigma T_{ab} t^a n^b d\Sigma
\label{CEAM}
\end{equation}
where
\begin{equation}
T_{ab} = \nabla_a\Phi \nabla_b \Phi
		- \frac{1}{2}g_{ab}(\nabla^c\Phi\nabla_c\Phi + m^2\Phi^2)
\label{stresstensor1}
\end{equation}
is the stress-energy tensor associated with the scalar field.

Ashtekar and Magnon define the quantum energy of the scalar field with respect
to $\Sigma$ (and $t^a$) to be the expectation value of the Hamiltonian operator
$H$, i.e.
\begin{equation}
QE_H = \left<\Phi,H\Phi\right>
\label{QEAM}
\end{equation}
But what is the Hamiltonian operator $H$?

If the vector field $t^a$ is Killing, so that the spacetime is stationary, the
usual definition for the Hamiltonian operator $H$ is
$H\Phi=-J(\pounds_{t^a}\Phi)$, where $\pounds$ represents Lie differentiation.
But in the present case the vector field $t^a$ is not necessarily Killing, and
so the function $\pounds_{t^a}\Phi$ is not necessarily a solution of the
Klein-Gordon equation.  Therefore $H$, as defined above, is not necessarily a
map into ${\cal H}$.

To overcome this problem, Ashtekar and Magnon used initial data to define $H$.
Let $\Phi\in V$ be a solution of the Klein-Gordon equation
with initial data as in (\ref{data}).  Consider the data to be a function of
$t$ and take its derivative; the result is in $\V$ and hence defines a
solution $\dot\Phi\in V$.  Explicitly, $\dot\Phi$ is the solution with initial
data
\footnote{The dot does not refer to a time derivative!  (This construction is
less intuitive with Ashtekar and Magnon's choice of data.)}
\begin{equation}
\pmatrix{\dot\varphi\cr \dot\pi}
  = \partial_t \pmatrix{\Phi\cr \sqrt{h}\,n^a\nabla_a\Phi} \bigg\vert_\Sigma
\label{AMdata2}
\end{equation}
It is straightforward but messy to use (\ref{KG}) to rewrite the time
derivatives in terms of spatial derivatives, resulting in
\begin{eqnarray}
\dot\varphi &=& {N\over\sqrt{h}} \,\pi \label{doteq1}\\
\dot\pi &=& \sqrt{h}\, (Nh^{ij}D_iD_j  + h^{ij}D_iND_j - m^2N) \,\varphi
\label{doteq2}
\end{eqnarray}
Ashtekar and Magnon proceed to define the Hamiltonian operator $H$ by
requiring
\begin{eqnarray}
H\Phi &=& -J\dot\Phi
\label{schrodee}
\end{eqnarray}

Using (\ref{CEAM}) and (\ref{QEAM}), the energy condition takes the form
\begin{equation}
\left<\Phi,H\Phi\right> = \int_\Sigma T_{ab} t^a n^b d\Sigma
\label{encond}
\end{equation}
It is again straightforward but messy to verify that
\begin{equation}
2\, \Omega(\Phi,\dot\Phi)
  = {\rm Re}\left<\Phi,H\Phi\right>
  = CE_T
\end{equation}
so that the true content of the energy condition is
\begin{equation}
2\, \Omega(\Phi,H\Phi)
  = {\rm Im} \left<\Phi,H\Phi\right>
  = 0
\label{true}
\end{equation}

We now state without proof Ashtekar and Magnon's main result.

\begin{theorem}
\label{AMtheo}
{\rm (Ashtekar and Magnon~\cite{AM})} Let $(M,g_{ab})$ be a globally
hyperbolic spacetime with Cauchy surface $\Sigma$, and let $V$ be as above.
Then there exists a unique compatible complex structure $J$ on $V$ such that
the energy condition is satisfied.  In other words, there is a unique complex
structure $J$ such that
\begin{equation}
\Omega(\Phi,H\Phi) = 0
\label{encond2}
\end{equation}
for every $\Phi\in\V$, where $H$ is defined in terms of $J$ via
(\ref{schrodee}).
\end{theorem}

It is important to note that Ashtekar and Magnon assume a particular form of
the Hamiltonian operator $H$, namely~(\ref{schrodee}).

%%%%%%%%%%%%%%%%%%%%%%%%%%%%%%%%%%%%%%%%%%%%%%%%%%%%%%%%%%%%%%%%%%%%%%%%%%%%%%
% Chapter 4

\section{Extending Ashtekar and Magnon's Result}
\label{Result}

The main result of this section is Theorem~\ref{theo1} which, is a
generalization of Theorem~\ref{AMtheo}.  There are two main differences.
First, we replace the energy condition with a more general condition, which
allows some flexibility in defining the classical energy of the scalar field.
Second, we eliminate the need for specifying the Hamiltonian operator $H$.
Theorem~\ref{theo1} uniquely determines both the complex structure and the
operator $\H$.

\begin{theorem}
\label{theo1}
Let $(M,g_{ab})$ be a globally hyperbolic spacetime with Cauchy surface
$\Sigma $.  Let $F$ be a real, nonzero smooth function on $\Sigma$, let $K$ be
the Cauchy completion of the inner product space $\tau$ with
$\ds\left<f,g\right>_\tau = \int_\Sigma fgF^{-1} \,d^{n-1}x$
and let $G$ be a real, semi-bounded, positive-definite symmetric operator on
$K$.  Suppose we have a linear operator $H$ and a compatible complex structure
$J$ defined on $V$ such that
\begin{equation}
\left<\Phi,H\Phi\right>
  = \frac{1}{2}\int_\Sigma (F\pi^2 + \varphi G\varphi)\,d^{n-1}x
\label{energycond}
\end{equation}
for all $\Phi\in V$ with data $\hat\Phi=\pmatrix{\varphi\cr \pi} \in \V$.
Then the operators $J$ and $H$ are unique and are given in terms of their
action on $\V$ by
\begin{eqnarray}
\J &=& \pmatrix{0& (FG)^{-\frac{1}{2}}F\cr -F^{-1}(FG)^{\frac{1}{2}}& 0}\\
\noalign{\medskip}
\H &=& \pmatrix{(FG)^{\frac{1}{2}}& 0\cr 0& F^{-1}(FG)^{\frac{1}{2}}F}
\end{eqnarray}
\end{theorem}

{\bf Proof 1:}
The right-hand-side of (\ref{energycond}) can be written as
\begin{equation}
\frac{1}{2} \Omega(\Phi,E\Phi) 
\end{equation}
where
\begin{equation}
\hat{E} = \pmatrix{~0& F\cr -G& 0}
\end{equation}
Comparing real and imaginary parts of (\ref{energycond}) yields for all
$\Phi\in V$:
\begin{eqnarray}
\Omega(\Phi,JH\phi) &=& \Omega(\Phi,E\Phi) \label{one}\\
\Omega(\Phi,H\Phi) &=& 0 \label{two}
\end{eqnarray}
and Lemma~\ref{symtheo2} now shows that both $H$ and $JH-E$ are symmetric.  As
discussed previously, $JH$ is antisymmetric if $H$ is symmetric, and
(\ref{firsteq})--(\ref{thirdeq}) (with appropriate sign changes) show that $E$
is antisymmetric.  Thus, $JH-E$ is both symmetric and antisymmetric, and we
conclude that
\begin{equation}
JH = E
\end{equation}
or equivalently
\begin{equation}
H = -JE
\label{hje}
\end{equation}
Using equation (\ref{hje}) we see that equation (\ref{two}) can be written as
\begin{equation}
\Omega(\Phi,JE\Phi) = 0
\label{jeiszero}
\end{equation}

Careful examination of the proof given by Ashtekar and Magnon shows that it
relies only on $J$ being a compatible complex structure and on
(\ref{jeiszero}).  We can thus use their proof to uniquely determine the
complex structure $\J$ in terms of the operators $F$ and $G$, the only
subtlety being the conditions on $G$ which allow square roots to be taken.
Finally, writing out the multiplication in (\ref{hje}) and using the identity
\begin{equation}
BG = -FC
\end{equation}
(which follows naturally from (\ref{jeiszero}) and the symmetry properties)
results in the given form for $\H$.  This completes the proof, full details of
which are given in~\cite{Jim}.
{\mbox{\endproof}}

It turns out there is another proof of Theorem~\ref{theo1}.  We provide this
alternate proof here:

{\bf Proof 2:}
Writing out the symmetry of $JH-E$ we obtain
\begin{eqnarray}
(AW + BY) & = & -(CX + DZ)^{\dag}\label{AXBZ}\\
((AX + BZ - F) & = & (AX + BZ - F)^{\dag}\label{CWBY}\\
((CW + DY + G) & = & (CW + DY + G)^{\dag}\label{CWDY}\\
X  & = &  -X^{\dag}\label{X}\\
Y  & = &  -Y^{\dag}\label{Y}\\
W  & = &  Z^{\dag}\label{ZW}
\end{eqnarray}
and the compatibility of $J$ yields
\begin{eqnarray}
B & = & B^{\dag}\label{B}\\
C & = & C^{\dag}\label{C}\\
A & = & -D^{\dag}\label{A=-D}\\
\end{eqnarray}
By using the symmetry and antisymmetry properties of the operators $A$ through
$Z$, we can rewrite (\ref{AXBZ})--(\ref{CWDY}) as
\begin{eqnarray}
AW + BY & = & XC + WA \label{XCWA}\\
XD + WB - F & = & -(AX + BZ - F)\label{XDWB}\\
ZC + YA + G & = & -(CW + DY + G)\label{ZCYA}
\end{eqnarray}
Taking the adjoint of (\ref{XCWA}) yields
\begin{equation}
ZD + YB = CX + DZ
\label{ZDYB}
\end{equation}
Solving for $F$ in (\ref{XDWB}) yields
\begin{equation}
F  =  \frac{1}{2}\left(AX + XD + WB + BZ\right)\label{F}
\end{equation}
Multiplying (\ref{F}) on the left by $A$ and on the right by $D$ and
subtracting gives
\begin{equation}
AF - FD  =  \frac{1}{2}\left(A^2X + AWB + ABZ - XD^2 - WBD -BZD\right)
\label{AFFD1}
\end{equation}
Multiplying (\ref{ZDYB}) on the left by $B$ and (\ref{XCWA}) on the right by
$B$ and solving for $BZD$ and $AWB$ yields
\begin{eqnarray}
BZD & = &  BCX + BDZ - BYB\label{BZD}\\
AWB & = & XCB + WAB - BYB\label{AWB}
\end{eqnarray}
Substituting (\ref{BZD}) and (\ref{AWB}) into (\ref{AFFD1}) yields
\begin{equation}
AF-FD = \frac{1}{2}\left(A^2X + BCX - XD^2 + ABZ - WBD + BDZ - WAB - XCB\right)\label{AFFD2}
\end{equation}

Finally, using (\ref{struct}) we see that the right hand side of (\ref{AFFD2})
is identically zero.  Therefore we now know that
\begin{equation}
D = F^{-1}AF\label{AFFD3}
\end{equation}
Using the argument given by Ashtekar and Magnon in proving
Theorem~\ref{AMtheo}, we can conclude that
\begin{equation}
A = 0 = D
\end{equation}
It is then straightforward to determine $B$ and $C$, thus obtaining the
complex structure $\J$, and to then solve for $\H$.
{\mbox{\endproof}}

%%%%%%%%%%%%%%%%%%%%%%%%%%%%%%%%%%%%%%%%%%%%%%%%%%%%%%%%%%%%%%%%%%%%%%%%%%%%%%
% Chapter 5

\section{Applications}
\label{Appl}

\subsection{Minimal Coupling ($ \xi = 0$)}
\label{application1}

We first recover Ashtekar and Magnon's result.  Comparing
(\ref{doteq1})--(\ref{doteq2}) and (\ref{schrodee}) with
(\ref{energycond}) shows that we should set
\begin{eqnarray}
G &=& \frac{\sqrt h}{~N} \,\Theta \\
F &=& \frac{~N}{\sqrt h}
\end{eqnarray}
where the operator $\Theta$ is defined by
\begin{equation}
\Theta = -\left( N^2 h^{ij} D_i D_j + h^{ij} N D_i N D_j - m^2 N^2 \right)
\end{equation}
Theorem~\ref{theo1} selects for us operators $\Jzero$ and $\Hzero$ (the zero
in the subscripts is being used to emphasize that we are considering the
minimally coupled case):
\begin{eqnarray}
\Jzero &=& \pmatrix{0& \Theta^{-\frac{1}{2}}\frac{~N}{\sqrt h}\cr
		\noalign{\smallskip}
		-\frac{\sqrt{h}}{~N}\Theta^{\frac{1}{2}}& 0}
\label{jay}\\
\noalign{\medskip}
\Hzero &=& \pmatrix{\Theta^{\frac{1}{2}}& 0\cr
	\noalign{\smallskip}
	0& \frac{\sqrt{h}}{~N}\Theta^{\frac{1}{2}}\frac{~N}{\sqrt{h}}}
\label{ache}
\end{eqnarray}
which agree with~\cite{AM}.

Furthermore, we have
\begin{equation}
\Jzero\Hzero\pmatrix{\varphi \cr \pi}
  = \pmatrix{\frac{~N}{\sqrt h}\pi\cr
	\noalign{\smallskip}
	 -\frac{\sqrt h}{~N}\Theta\varphi}
\end{equation}
and comparing with (\ref{doteq1})--(\ref{doteq2}) shows that
\begin{eqnarray}
\hat\Hzero\hat\Phi &=& -\Jzero\hat{\dot\Phi}
\end{eqnarray}
as desired.  Therefore the Hamiltonian operator $H$ and complex structure $J$
obtained via Theorem~\ref{theo1} satisfy an equation which mimics the
Schr\"odinger equation.  It is of course not always true that $\dot\Phi =
t^a\nabla_a\Phi$.  However, if the vector field $t^a$ is Killing, then the
operators $H$ and $J$ determined by Theorem~\ref{theo1} would indeed satisfy
{\bf the} Schr\"odinger equation
\begin{eqnarray}
H\Phi & = & -J\pounds_t\Phi \label{shrodenew}
\end{eqnarray}
and would therefore correctly reduce to the well-established theory for static
spacetimes.

\subsection{Non-Minimal Coupling ($\xi \neq 0$)}
\label{application2}

For the second application of Theorem~\ref{theo1} we will allow non-zero
values for the coupling constant $\xi$.  As in the previous application we
will need to define what is meant by the classical energy of the scalar field.
If we stick with the definition for the classical energy which is given by
(\ref{CEAM}) we will find that Theorem~\ref{theo1} selects for us a
Hamiltonian operator and complex structure.  However, it turns out that these
operators do not satisfy (\ref{schrodee}) and hence, in the static limit, the
operators would not satisfy (\ref{shrodenew}).  However, by choosing the
classical energy of the field to be the surface integral of the classical
Hamiltonian, we can still apply Theorem~\ref{theo1}, and in this case we
obtain a Hamiltonian operator and complex structure which do reduce to the
usual Hamiltonian operator and complex structure when considering static
spacetimes.

A primary candidate for our definition of the classical energy of the field
associated with the Cauchy surface $\Sigma$ and timelike vector field $t^a$ is
the one given by Ashtekar and Magnon (\ref{CEAM}), involving the stress-energy
tensor $T_{ab}$.  The stress-energy tensor is is obtained by varying the
action (\ref{action}) with respect to the metric $g_{ab}$ (for more details
see~\cite{BD}, Chapter~3).
\footnote{There is a sign error in the last term of the third equation of
(3.196) on p.\ 88 in~\cite{BD}.}
\begin{eqnarray}
 T_{ab}= \frac{2}{\sqrt{-g}}\frac{\delta S}{\delta g^{ab}} & = & (1 - 2\xi
)\nabla _a \Phi \nabla _b \Phi + (2\xi - \frac{ 1}{ 2})g_{ab}g^{cd}\nabla
_c\Phi \nabla _d \Phi \nonumber\\ &  & \mbox{} - 2\xi \Phi \nabla _a \nabla _b
\Phi + 2\xi g_{ab}\Phi g^{cd}\nabla_c\nabla_d\Phi \Phi \nonumber\\ &  &
\mbox{} + \frac{1}{2}( g_{ab}m^2 +  g_{ab}R\xi - 2\xi R_{ab} ) \phi ^2
\label{Stress}
\end{eqnarray}
In the minimally coupled case, that is when $\xi = 0$, the stress-energy
tensor reduces to (\ref{stresstensor1}).  Straightforward computations allow
us to put the integral $\int_\Sigma T_{ab}n^a t^b d\Sigma$ in the following
form:
\begin{equation}
\int _\Sigma T_{ab}n^a n^b Nd\Sigma
  = \frac{1}{2} \int _\Sigma \left(
	\frac{~N}{\sqrt h}\pi ^2 + \frac{\sqrt h}{~N} \varphi \Gamma \varphi 
		\right)\,  d^{n-1}x
\end{equation}
where
\begin{eqnarray}
-\Gamma & = & (1 - 4\xi)h^{ij} N D_i N D_j + N^2 (1 - 4\xi) h^{ij} D_i D_j \\
        &   & \mbox{ } - \xi R N^2  - 2\xi n^an^bR_{ab} - m^2 N^2
\end{eqnarray}
Since $\Gamma$ is semi-bounded, symmetric and positive-definite on $K$ and
$\frac {~N}{\sqrt h}$ is nonzero, we can apply Theorem~\ref{theo1} and obtain
the following operators:

\begin{eqnarray}
 \J_{\xi} & = & \pmatrix {0 & \Gamma^{-\frac{1}{2}}\frac{~N}{\sqrt h} \cr
-\frac{{\sqrt h}}{~N}\Gamma ^{\frac{1}{2}} & 0}\label{jay2}\\
& &\nonumber \\
 \H_{\xi} & = & \pmatrix{\Gamma^{\frac{1}{2}} & 0 \cr
	0 & \frac{{\sqrt h}}{~N}\Gamma^{\frac{1}{2}}\frac{~N}{\sqrt h}}
\label{ache2}
\end{eqnarray}

We can proceed as we did in the minimally coupled case and find that the
operators (\ref{jay2}) and (\ref{ache2}) satisfy
\begin{equation}
\Hxi\pmatrix{\varphi \cr \pi}
  = -\Jxi\pmatrix{{\frac{~N}{\sqrt h}\pi}\cr
	\noalign{\smallskip}
	-\frac{\sqrt h}{~N}\Gamma\varphi}
\label{shrode3}
\end{equation}
As in the minimally coupled case we have that
\begin{equation}
\frac{~N}{\sqrt h}\pi = \partial_t\Phi\vert_\Sigma\nonumber
\end{equation}
However, the function $-\frac{\sqrt h}{~N}\Gamma\varphi$ does not equal the
restricted time derivative of $\Pi=\sqrt{h}n^a\nabla_a\Phi$:
\begin{eqnarray}
-\frac{\sqrt h}{~N}\Gamma\varphi
  & = & \sqrt h(1 - 4\xi)h^{ij}D_iND_j\varphi
	+ N(1 - 4\xi)h^{ij}D_i\varphi D_j\varphi \nonumber \\
  & & \mbox{ } - (\xi RN + N^{-1}2\xi n^an^bR_{ab} + m^2N)\varphi
\label{Psi}\\
	\partial_t\Pi\vert_\Sigma & = & \sqrt hh^{ij}D_iND_j\varphi
	+ Nh^{ij}D_i\varphi D_j\varphi \nonumber \\
  &   & \mbox{ } - (\xi RN + m^2N)\varphi
\label{Pidot}
\end{eqnarray}
There are several differences between (\ref{Psi}) and (\ref{Pidot}), so that
(\ref{shrode3}) does not mimic the Schr\"odinger equation.  We conclude that
using the stress-energy tensor to define the classical energy of the field
when $\xi \neq 0$ will produce an undesirable choice for the Hamiltonian
operator and complex structure.

If $\xi \neq 0$, one is either forced to reconsider the definition of the
classical energy of the scalar field or abandon the use of
Theorem~\ref{theo1}.  Fortunately there does exist at least one other natural
method for defining the classical energy of the field; this alternate
definition involves the classical Hamiltonian.

With $\Phi$ and $\Pi$ as above, the classical Hamiltonian
\footnote{The usual definition for the classical Hamiltonian assumes that
$t^a$ is a Killing vector field~\cite{SF,JR}.  Therefore it may be more
appropriate to call this a generalized classical Hamiltonian.}
is defined to be
\begin{eqnarray}
\Ham & = & \Pi t^a\nabla_a\Phi - {\cal L}\\
     & = & \frac{~N}{\sqrt h}\Pi^2 - {\cal L}\\
     & = & \frac{1}{2}\left(\frac{~N}{\sqrt h}\Pi^2 + N{\sqrt h}
	h^{ij}\partial_i\Phi\partial_j\Phi + N\sqrt h(m^2 + \xi R)\Phi^2\right)
\label{Ham}
\end{eqnarray}
The alternate definition for the classical energy of the scalar field
associated to the hypersurface $\Sigma$ and the vector field $t^a$ is the
surface integral of this Hamiltonian:
\begin{equation}
\CE2 = \int_\Sigma \Ham \,d^{n-1}x\label{Hamdensity}\\
     = \frac{1}{2}\int_\Sigma \left(\frac{~N}{\sqrt h}\pi^2
	+ N{\sqrt h}h^{ij}D_i\varphi D_j\varphi
	+ N{\sqrt h}(m^2 + \xi R)\varphi^2 \right)\,d^{n-1}x
\label{Hamdensity2}
\end{equation}
In the case of minimal coupling, this definition for the classical energy of
the scalar field coincides with Ashtekar and Magnon's definition which
involves the stress-energy tensor.  We now show that by using the surface
integral of the classical Hamiltonian to represent the classical energy of the
field, we are still able to apply Theorem~\ref{theo1}.  Moreover,
Theorem~\ref{theo1} determines a unique Hamiltonian operator and unique
complex structure that reduce to the appropriate operators when considering
static spacetimes.

By using the definition of the Hamiltonian and applying integration by parts
we obtain
\begin{equation}
\CE2 = \frac{1}{2}\int_\Sigma \left(
	\frac{~N}{\sqrt h}\pi^2
	-\varphi ({\sqrt h}(Nh^{ij}D_iD_j
	+ h^{ij}D_iND_j - N (m^2 + R\xi))\varphi
		\right)\, d^{n-1}x
\end{equation}
The operator
\begin{equation}
-\Upsilon = N^2 h^{ij} D_i D_j + h^{ij} N D_i N D_j - N^2 (m^2 + R\xi)
\end{equation}
is positive-definite, semi-bounded and symmetric on $K$.  We can therefore
apply Theorem~\ref{theo1} by letting $G = \frac{\sqrt h}{~N}\Upsilon$ and
$F = \frac{~N}{\sqrt h}$.  We obtain for $\Jxi$ and $\Hxi$ the following
operator-valued matrices:
\begin{eqnarray}
 \J_{\xi} & = & \pmatrix {0 & \Upsilon^{-\frac{1}{2}}\frac{~N}{\sqrt h} \cr
		-\frac{{\sqrt h}}{~N}\Upsilon ^{\frac{1}{2}} & 0}
\label{jay3}\\
\noalign{\medskip}
 \H_{\xi} & = & \pmatrix{\Upsilon^{\frac{1}{2}}& 0\cr 
	0& \frac{{\sqrt h}}{~N}\Upsilon^{\frac{1}{2}}\frac{~N}{\sqrt h}}
\label{ache3}
\end{eqnarray}

Proceeding as in the previous cases, we find that the operators (\ref{jay3})
and (\ref{ache3}) satisfy
\begin{equation}
\Hxi\pmatrix{\varphi \cr \pi} = -\Jxi\pmatrix{{\frac{~N}{\sqrt h}\pi} \cr
	-\frac{\sqrt h}{~N}\Upsilon\varphi}
\label{shrode4}
\end{equation}
However, this time we have
\begin{equation}
\pmatrix{{\frac{~N}{\sqrt h}\pi} \cr -\frac{\sqrt h}{~N}\Upsilon\varphi}
  = \partial_t \pmatrix{\Phi\cr \sqrt{h}\,n^a\nabla_a\Phi} \bigg\vert_\Sigma
\end{equation}
That is, (\ref{shrode4}) does indeed reduce to the Schr\"odinger equation when
$t^a$ is a Killing vector field.

%%%%%%%%%%%%%%%%%%%%%%%%%%%%%%%%%%%%%%%%%%%%%%%%%%%%%%%%%%%%%%%%%%%%%%%%%%%%%%
% Chapter 6

\section{Discussion}
\label{Disc}

In Section~\ref{Energy} we summarized Ashtekar and Magnon's uniqueness result
(Theorem~\ref{AMtheo}).  By requiring the quantum energy of the scalar field
to be equal to the classical energy of the field at each instant of time, they
were able to uniquely specify a complex structure $\J$ at each instant of
time.  However, their result depended on the need to define the Hamiltonian
operator.

Our main result (Theorem~\ref{theo1}) was discussed in Section~\ref{Result}.
We showed that the Ashtekar and Magnon energy condition uniquely determines
not only the complex structure, but also the Hamiltonian operator.  As shown
in Section~\ref{application1}, Ashtekar and Magnon's result is thus a special
case of Theorem~\ref{theo1}.

An important consequence of Theorem~\ref{theo1} concerns the case of
non-trivial coupling ($\xi\neq0$).  We saw in Section~\ref{application2} that
the usual definition for the classical energy produces a complex structure and
Hamiltonian operator that do not reduce to the appropriate operators if the
spacetime is assumed to be static.  However, also in
Section~\ref{application2}, we showed that using the classical Hamiltonian to
define the classical energy produces via Theorem~\ref{theo1} a complex
structure and Hamiltonian operator which do have the correct limits in the
static case.

It is somewhat disturbing that the two obvious formulations of the classical
energy, namely the stress-energy tensor and the classical Hamiltonian, fail to
agree when $\xi\ne0$.  The results of Section~\ref{application2} suggest that
the latter is to be preferred.

It would therefore be worthwhile to further examine the properties of the
generalized classical energy (\ref{Ham}).  For instance, under what
circumstances is it conserved?  The stress-energy tensor (\ref{Stress}) is
obtained by varying the action (\ref{action}) with respect to the metric
$g_{ab}$.  By suitably modifying the action prior to carrying out the
variation, is it possible to obtain the same classical energy using the
stress-energy tensor that is obtained using the classical Hamiltonian?  We
have shown that if $\xi\ne0$ then the complex structure and Hamiltonian
operator obtained using the stress-energy tensor are different from those
obtained using the classical Hamiltonian.  What is the relationship between
the two different Fock spaces which are associated with these two different
choices for the classical energy?

Finally, we emphasize that both Ashtekar and Magnon's work and ours consider
only hypersurface orthogonal observers.  While this description lends itself
naturally to globally hyperbolic spacetimes, in which one can always {\it
choose} such observers, it does not address stationary but non-static
observers, let alone more general rotating observers.  Some preliminary ideas
on how to deal with these cases can be found in \cite{Kulkarni,Boersma}.

\section*{Acknowledgments}

It is a pleasure to acknowledge discussions with Corinne Manogue.  This work
is based on a dissertation~\cite{Jim} submitted to Oregon State University by
JF in partial fulfillment of the requirements for the Ph.D.\ degree in
mathematics.

%%%%%%%%%%%%%%%%%%%%%%%%%%%%%%%%%%%%%%%%%%%%%%%%%%%%%%%%%%%%%%%%%%%%%%%%%%%%%%
% References

%%%%%%%%%%%%%%%%%%%%%%%%%%%%%%%%%%%%%%%%%%%%%%%%%%%%%%%%%%%%%%%%%%%%%%%%%%%%%%

\end{document}